\documentclass{elsart3}
\usepackage{epsfig}
\begin{document}

\begin{frontmatter}

\title{Giant Quantum Oscillations of the Longitudinal Magnetoresistance in Quasi two-dimensional Metals}

\author{T. Champel and V.P. Mineev}
\address{
Commissariat \`{a} l'Energie Atomique,  DSM/DRFMC/SPSMS\\
17 rue des Martyrs, 38054 Grenoble Cedex 9, France}
\date{\today}

\begin{abstract}
We have investigated in frame of the quantum transport theory the magnetic quantum oscillations of the longitudinal magnetoresistance $\rho_{zz}$ in quasi two-dimensional metals for a magnetic field perpendicular to the layers. 
 Giant Shubnikov-de Haas oscillations are found when the cyclotron energy $\hbar \omega_{c}$ is much larger than the interlayer transfer integral $t$ (the two-dimensional limit). In large magnetic fields and at low temperatures, the minima of the magnetoconductivity $\sigma_{zz}=\rho_{zz}^{-1}$ exhibit a thermally activated behavior in presence of negligibly small chemical potential oscillations, as observed in the organic layered conductor  $\beta''\mathrm{-(BEDT-TTF)}_{2}\mathrm{SF}_{5}\mathrm{CH}_{2}\mathrm{CF}_{2}\mathrm{SO}_{3}$. 
 The questions concerning the absence of strong chemical potential oscillations in such compound and the impurity self-energy are discussed.

\end{abstract}


\end{frontmatter}

\section{Introduction}

Features of the magnetic quantum oscillations of magnetization (de Haas-van Alphen effect) or of magnetoresistance (Shubnikov-de Haas effect) are known to be noticeably different in two-dimensional (2D) metals from three-dimensional (3D) metals. For example, the 2D de Haas-van Alphen effect is characterized by a sharp sawtooth-like shape at low temperatures, while the oscillations of the magnetization in 3D metals are always smooth.
The difference between 2D and 3D metals can be principally understood  in terms of Landau levels broadening.

In the 3D case, the Fermi energy is crossed by many occupied Landau levels due to the presence of states in the direction parallel to the magnetic field $H$. These non quantized states available for any amplitude of magnetic field act as a reservoir. 
The discrete nature of the Landau levels is thus blurred by this broadening inherent to the dimensionality of the spectrum.
As a result, the chemical potential $\mu$ oscillates negligibly with the magnetic field.
Moreover,
the oscillating part of the transport properties is always small compared to the non-oscillating part. 
For example, in a metal with a spherical Fermi surface, the oscillating part $\Delta \sigma$ of the conductivity $\sigma$ is of the order of~\cite{Abr}

\begin{equation}
\frac{\Delta \sigma}{\sigma} \sim 
\left(\frac{\omega_{c}}{\mu}\right)^{2}
\frac{\partial M_{\mathrm{osc}}}{\partial H}
\sim\sqrt{\frac{\omega_{c}}{\mu}} \ll 1,
\end{equation}
where $\omega_{c}$ is the cyclotron pulsation, and $M_{\mathrm{osc}}$ is the oscillating part of magnetization.

In the 2D case, there is no Landau levels broadening due to dimensionality, and thus the oscillations of the thermodynamic and transport properties can be very strong. The principal source of Landau level broadening is henceforth the presence of impurities in real materials. 
In 3D metals, the impurity broadening is an accessory ingredient in the derivation of theory of magnetic quantum oscillations, and is usually taken into account by hand by introducing in the $l$th harmonic amplitude of the oscillations
 the so-called Dingle factor $$\exp(-2 \pi l \Gamma_{0}/\omega_{c})$$
 with $\Gamma_{0}$ the zero field value of the imaginary part of the impurity self-energy. Such a factor has received a microscopic justification in Ref.~\cite{Byc}.
On the other hand, the consideration of impurities effects on the magnetic quantum oscillations is of the first importance in the theory of magnetic quantum oscillations in 2D metals.
The broadening due to point-like impurities is usually investigated in the 
so-called self-consistent Born approximation (SCBA) which leads to oscillations of the imaginary part of the energy-dependent impurity self-energy $\Gamma(\varepsilon)$. In high magnetic fields, the oscillations of $\Gamma(\varepsilon)$ become very strong. In such a regime, 
the mere consideration of point-like impurities has been shown to be inappropriate to describe correctly the magnetic quantum oscillations~\cite{Rai,Lai}. To our knowledge, the complete and satisfactory theory of the impurities effects on magnetic quantum oscillations, i.e. of the lifting of the degeneracy of the Landau levels due to impurities, still does not exist.

Another problem which arises in 2D metals is the presence of nonnegligible chemical potential oscillations when the system is isolated. Actually, the influence of oscillations of $\mu$ on the shape of the magnetization oscillations or magnetoresistivity oscillations can be investigated at a semi-phenomenological level (by bypassing the microscopic problem with the impurities which determines the exact dependence of the chemical potential with the magnetic field) and  makes no important difficulty~\cite{Cha2001b}.

In quasi-2D metals with the magnetic field perpendicular to the layers, the energy spectrum has the form
\begin{equation}
\varepsilon_{n}(p_{z})=(n+1/2) \omega_{c}-2t\cos p_{z}s
\label{Spectreq2D}
\end{equation}
with $p_{z}$ the interlayer momentum, $t$ the interlayer transfer integral and $s$ the interlayer distance.
The situation concerning the Landau levels broadening is in some sense intermediate between the 2D and 3D cases and depends on the ratio $t/\omega_{c}$  : the Landau levels broadening due to dimensionality is strongly effective when $t \gg \omega_{c}$. In the opposite limit, when $t \ll \omega_{c}$, the Landau levels are broadened only through the impurities effects, exactly 
like in 2D metals.
It is worth noting that the 3D de Haas-van Alphen and Shubnikov-de Haas theories are not directly applicable to quasi-2D metals because the saddle-point method used in the calculations of the 3D case~\cite{Abr} breaks down with the spectrum (\ref{Spectreq2D}) when $\omega_{c} \geq t$.
The (semi-phenomenological) theory of the de Haas-van Alphen effect in quasi-2D metals~\cite{Cha2001a} has revealed a crossover between the 2D and 3D limits for the behavior of the magnetization oscillations.

It was with the same idea of a reduced dimensionality driven by the ratio $t/\omega_{c}$ that we studied the magnetoresistance oscillations in quasi-2D metals~\cite{Cha2002}. The regime $t \geq \omega_{c}$ has been investigated in details in Ref.~\cite{Gri2003} and in references therein. 
Here, we are rather interested in the 2D limit $t \ll \omega_{c}$, motivated by the experiments of Wosnitza {\em et al.}~\cite{Wos2001} and  Nam {\em et al.}~\cite{Nam2001} on the organic layered conductor $\beta''\mathrm{-(BEDT-TTF)}_{2}\mathrm{SF}_{5}\mathrm{CH}_{2}\mathrm{CF}_{2}\mathrm{SO}_{3}$. Giant Shubnikov-de Haas oscillations of the longitudinal magnetoresistance $\rho_{zz}$ were observed at high magnetic fields and low temperatures. In particular, a regime with a thermally activated behavior of the minima of the magnetoconductivity $\sigma_{zz}=\rho_{zz}^{-1}$ has been pointed out for magnetic fields higher than 20 T~\cite{Nam2001}. Such a feature is completely absent in the 3D Shubnikov-de Haas theory.

Strong oscillations of $\rho_{zz}$ were also reported in the organic conductor $\alpha-(\mathrm{ET})_{2}\mathrm{TlHg}(\mathrm{SeCN})_{4}$ \cite{Lau1995}. The observation of slow oscillations of the magnetoresistance in a field region \cite{Lau1995} implies that the situation in this compound is presumably intermediate between the two regimes $t \geq \omega_{c}$ \cite{Gri2003} and $t \ll \omega_{c}$ \cite{Cha2002}.

As mentioned previously, the 2D limit  in quasi-2D metals is problematic concerning the theoretical treatment of the impurities effects. Being aware of this weakness, only qualitative properties based on general theoretical considerations (i.e. without making specific assumptions about the form of the impurity self-energy) are expected to be reasonable at this stage.

The relevance of the 2D limit in the conductor $\beta''\mathrm{-(BEDT-TTF)}_{2}\mathrm{SF}_{5}\mathrm{CH}_{2}\mathrm{CF}_{2}\mathrm{SO}_{3}$ is indicated by the sawtooth shape of the de Haas-van Alphen signal  already observed for fields $H \geq 10 $ T~\cite{Wos2000}.
Moreover, the magnetization ocillations~\cite{Wos2000} are also consistent with negligibly small chemical potential oscillations in this compound. The reason for this is still unclear. An explanation may be the presence of an intrinsic  non-quantized reservoir formed by quasi-1D bands. We note that another mechanism for reducing the chemical potential oscillations may be the magnetostriction. Indeed, the volume of the sample is not  fixed in the conditions of experiments and thus is expected to vary with the variation of magnetization. The changes in the chemical potential associated with the oscillations of the density of states 
and with the magnetostriction might be cancelled under particular conditions~\cite{Ale}.

In the following sections, we derive first the form
of $\sigma_{zz}$ in the 2D limit for a point-like impurity scattering with an arbitrary impurity self-energy. The mechanism of appearance of the thermal activation of conductivity minima is pointed out. Then, assuming a specific form for the impurity self-energy, we have performed a numerical analysis of the temperature and magnetic field dependences of $\rho_{zz}$ oscillations to compare with experiments~\cite{Wos2001,Nam2001}.

\section{Quantum oscillations of $\sigma_{zz}$}
We restrict ourselves to point-like impurity scattering, which allows us to neglect the vertex corrections.
The magnetoconductivity $\sigma_{zz}$ consists of two pieces susceptible of oscillating with the magnetic field~\cite{Cha2002}
\begin{equation}
\sigma_{zz}=\int_{-\infty}^{+\infty}d \varepsilon \left(-f'(\varepsilon) \right)\left[\sigma_{RA}(\varepsilon)+\sigma_{RR}(\varepsilon)\right],
\end{equation}
the Boltzmann piece 
\begin{eqnarray}
\sigma_{RA}(\varepsilon)=e^{2} g_{0} \frac{\omega_{c}}{2\pi}\sum_{n}\int\frac{dp_{z}}{2\pi}v_{z}^{2}(p_{z}) 
G^{R}_{n,p_{z}}G^{A}_{n,p_{z}},
\end{eqnarray}
and a pure quantum piece 
\begin{eqnarray}
\sigma_{RR}(\varepsilon)=-e^{2} g_{0} \frac{\omega_{c}}{2\pi}\sum_{n}\int\frac{dp_{z}}{2\pi}v_{z}^{2}(p_{z}) 
\Re \, \left(G^{R}_{n,p_{z}}\right)^{2}.
\end{eqnarray}
Here
$f'(\varepsilon)$ is the derivative of the Fermi Dirac distribution function, $g_{0}$ is the constant 2D density of states (at $H=0$),
\begin{equation}
v_{z}(p_{z})=\frac{\partial \varepsilon_{n}(p_{z})}{\partial p_{z}}=2st\sin p_{z}s,
\end{equation}
and
\begin{equation}
G_{n,p_{z}}^{A}=\left(G_{n,p_{z}}^{R}\right)^{\ast}=\frac{1}{\varepsilon-\varepsilon_{n}(p_{z})+ i \Gamma(\varepsilon)}
\end{equation}
 are the advanced and retarded Green's functions.

In the 2D limit $\omega_{c} \gg t$, the summation over the Landau levels can be performed analytically to give~\cite{Cha2002}

\begin{equation}
\sigma_{RA}(\varepsilon)=\sigma_{0}\chi_{0}
\frac{
\sinh\chi(\varepsilon)/\chi(\varepsilon)}
{\cosh\chi(\varepsilon)+
\cos\left(2 \pi \frac{\varepsilon}{\omega_{c}}\right)}
\label{RA}
\end{equation}
and
\begin{equation}
\sigma_{RR}(\varepsilon)=-\sigma_{0}
\chi_{0}
\frac{
1+\cosh\chi(\varepsilon)\cos\left(2 \pi \frac{\varepsilon}{\omega_{c}}\right)}
{\left(\cosh\chi(\varepsilon)+
\cos\left(2 \pi \frac{\varepsilon}{\omega_{c}}\right)\right)^{2}}
\label{RR}
\end{equation}
with 
$$
\sigma_{0}=e^{2}g_{0}t^{2}s/\Gamma_{0}, \hspace{0.5cm}
\chi_{0} =2 \pi \frac{\Gamma_{0}}{\omega_{c}}, \hspace{0.5cm} \chi(\varepsilon)=2 \pi \frac{\Gamma(\varepsilon)}{\omega_{c}}.
$$
Here the function $\chi(\epsilon)$ is still unknown, since no specific assumption about the impurity self-energy has been made yet. 

\begin{figure}
\epsfig{file=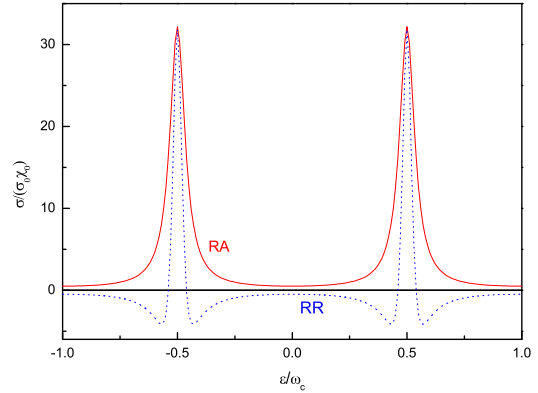,height=6.5cm}
\caption{Both contributions $\sigma_{RA}$ (solid line) and $\sigma_{RR}$ (dotted line) for a constant parameter $\chi(\varepsilon)=0.25$. They compensate each other between the Landau levels and add together at the Landau levels.} 
\label{fig1}
\end{figure}

For $\chi(\varepsilon) \ll 1$, it is straightforward to note that 
the two contributions $\sigma_{RA}$ and $\sigma_{RR}$ are compensated between the Landau levels, and are summed at the Landau levels, to give 
\begin{equation}
\sigma_{zz}(\varepsilon)=\frac{4}{3}\sigma_{0}\chi_{0}
\frac{
\chi^{2}(\varepsilon)
\left(1+2\sin^{2}\left(\pi \frac{\varepsilon}{\omega_{c}}\right)
\right)
}
{\left(\chi^{2}(\varepsilon)+4\cos^{2}\left(\pi \frac{\varepsilon}{\omega_{c}}\right)\right)^{2}}.
\end{equation}
The spectral conductivity $\sigma_{zz}(\varepsilon)$ between the Landau levels is very small and flat over a wide range in energy. 
To illustrate the particular behavior of $\sigma_{zz}(\varepsilon)$ at $\chi(\varepsilon) \ll 1$, we have  represented in  Fig. \ref{fig1} the functions (\ref{RA}) and (\ref{RR}) for a constant $\chi(\varepsilon)=0.25$.

\section{Maxima of magnetoresistivity}
If the chemical potential oscillations are negligibly small, the main means of conduction between the Landau levels at a finite low temperature is the thermal excitation at the edges of the gap-like range $\Delta$. This explains why the conductivity is minimal for integer values of $\mu/\omega_{c}$ with the thermally activated dependence 
\begin{equation}
\sigma_{zz}^{\mathrm{min}} \sim \exp\left(-\frac{\Delta}{T}\right).
\end{equation}
At very low temperatures, quasi-particles whose energy lies in the $\Delta$ range mainly contribute to conduction, and the minima of conductivity saturate.

If the thermally activated behavior can be pointed out by qualitative arguments, i.e. without invoking a specific model for $\Gamma(\varepsilon)$, the study at a quantitative level needs
specific assumptions concerning the impurity self-energy.
As declared in introduction, such investigation has to be done with care, since the theory in the 2D limit does not exist yet.

\begin{figure}
\epsfig{file=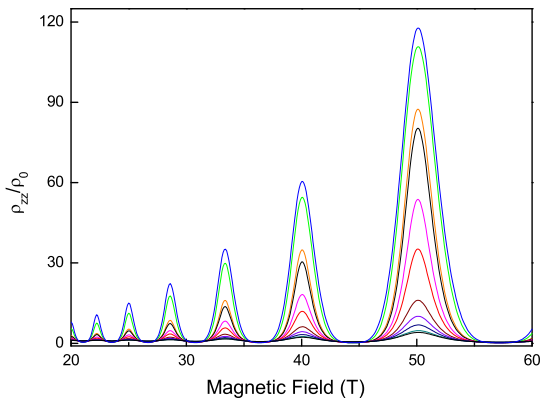,height=6.5cm}
\caption{Oscillations of the magnetoresistance $\rho_{zz}/\rho_{0}=\sigma_{0}/\sigma_{zz}$ as a function of magnetic field for different temperatures (from the top, 0.59, 0.94, 1.48, 1.58, 1.91, 2.18, 2.68, 3.03, 3.38, 3.80, and 4.00 K).} 
\label{fig2}
\end{figure}

Since the chemical potential may be fixed in  $\beta''\mathrm{-(BEDT-TTF)}_{2}\mathrm{SF}_{5}\mathrm{CH}_{2}\mathrm{CF}_{2}\mathrm{SO}_{3}$ by a finite reservoir formed by  a quasi-1D band~\cite{Wos2000}, we proposed~\cite{Cha2002} to include this reservoir in the calculations of $\Gamma(\varepsilon)$ within a model of short-range impurity scattering. Using the 
SCBA, we obtained the self-consistent equation~\cite{Cha2002}
\begin{equation}
\chi(\varepsilon)=\frac{\chi_{0}}{1+R}\left[R+
\frac{
\sinh\chi(\varepsilon)}
{\cosh\chi(\varepsilon)+
\cos\left(2 \pi \frac{\varepsilon}{\omega_{c}}\right)}\right]
\label{SCBA}
\end{equation}
where $R$ is a dimensionless parameter measuring the strength of the finite reservoir. For $R=0$ (no reservoir) and $\chi_{0} \leq 2$, this equation leads to the absence of states between the Landau levels. This unphysical result~\cite{Rai,Lai} is avoided as soon as $R\neq 0$. Assuming the value $R=5$ necessary for suppressing significantly the chemical potential oscillations~\cite{Wos2000}, we found~\cite{Cha2002} that the oscillations of $\chi(\varepsilon)$ are of the order of 10 \% for $H=60$ T, and thus neglected them in our numerical calculation~\cite{Cha2002} (i.e. we took $\chi(\varepsilon) \approx \chi_{0}$).
In Fig. \ref{fig2}, we show the oscillations of $\rho_{zz}$ at different temperatures. The thermal activation of $\rho_{zz}$ maxima is well reproduced numerically for fields $H \geq 20$ T.

\section{Concluding remarks}
Our model considering only short-range impurity scattering and the spectrum (\ref{Spectreq2D}) therefore accounts for the thermal activation of $\sigma_{zz}$ minima  in presence of negligibly small chemical potential oscillations,
as observed  in the layered conductor $\beta''\mathrm{-(BEDT-TTF)}_{2}\mathrm{SF}_{5}\mathrm{CH}_{2}\mathrm{CF}_{2}\mathrm{SO}_{3}$. However, there remain some qualitative features in experiments~\cite{Wos2001,Nam2001} not explained by our theoretical model, such as the increase of $\rho_{zz}$ minima with the magnetic field at high magnetic fields and low temperatures.
It is worth noting that under these latter conditions, 
the Landau levels broadening due to point-like impurity scattering alone is probably ineffective. Landau levels broadening due to
impurity scattering on finite range becomes predominant and has to be envisaged as being physically relevant for seizing the complete behavior of $\rho_{zz}$ oscillations.

\end{document}